\documentclass[doublecol]{epl2} 
\title{Mutually unbiased measurements for high-dimensional time-bin based photonic states}
\shorttitle{Mutually unbiased measurements} %Insert here a short version of the title if it exceeds 70 characters

\author{Thomas Brougham  \and Stephen M. Barnett}
\shortauthor{T. Brougham and S. M. Barnett}

\institute{School of Physics and Astronomy, University of Glasgow, Glasgow, G12 8QQ, U.K.
}

\abstract{
The task of measuring in two mutually unbiased bases is central to many quantum information protocols, as well as being of fundamental interest.  Increasingly, there is an experimental focus on generating and controlling high-dimensional photonic states.  One approach is to use the arrival time of a photon, which can be split into discrete time bins.  An important problem associated with such states is the difficulty in experimentally realizing a measurement that is mutually unbiased with respect to the time-of-arrival.  We propose a simple and compact scheme to measure in both the time of arrival basis and a basis that is approximately mutually unbiased with respect to the arrival time.%This scheme has both fundamental interest as well as having applications within quantum information e.g. in high-dimensional quantum key distribution.}
}
\begin{document}

\maketitle

{\bf Introduction}. - One of the most puzzling aspects of quantum mechanics is that observable quantities can be complementary to one another.  This fact is illustrated by the famous commutation relation for position and momentum.  For finite dimensional systems, the notion of complementarity is captured in the concept of mutually unbiased bases (MUBs) \cite{schwinger,wootters}.  For a given $d$-level system, two orthonormal bases $\{|a_m\rangle\}$ and $\{|b_n\rangle\}$, are said to be mutually unbiased if and only if
\begin{equation}
\label{mubcond}
|\langle a_m|b_n\rangle|^2=1/d,
\end{equation}
for each element of the two bases.  One can think of these two bases as being the eigen-vector of two conjugate observables, $\hat A$ and $\hat B$\cite{pegg90}.  The relation (\ref{mubcond}) shows that knowledge of one of these observables implies a complete lack of information about the other.  

Mutually unbiased bases play a fundamental role within quantum physics.  For example, if one wants to maximise the violation of local realism, then the best choice of measurement bases for each party will often be ones that are mutually unbiased \cite{bellviolation}.  Another application is state estimation, where one aims to determine an unknown state by making suitable measurements within several different bases.  Inevitably, any estimate of the state will be imperfect.  It has been shown that one can minimize the error in the estimate by measuring in bases that are mutually unbiased to one another \cite{wootters}.  Quantum key distribution provides another application where it is essential to be able to measure within at least two MUBs \cite{bennettbrassard,qkdreview,NPWBK,NRA,CBKG,barnett}.

Given the many applications of MUBs, it is important to be able to measure experimentally in a pair of such bases.  If our system is a qubit, then this task can often be achieved straightforwardly.  For example, the qubit could be realized in terms of the polarization degrees of freedom of a photon.  One could thus measure in a pair of MUBs by using two polarizing beam splitters.  The task of measuring within two MUBs can, however, becomes much more involved when one considers high-dimensional systems.

A common approach to generating high-dimensional optical states, is to make use of the arrival time of a photon.  The possible arrival times can be split into discrete time slots or time bins.  Using this approach one can engineer states that live in very high-dimensional Hilbert spaces.  This approach has been used to generate photon pairs that are entangled within their time of arrival, so called energy-time entanglement \cite{tb1,tb2,tb3,tb4,BLPK}.  Such states have been used to demonstrate non-local effects within time \cite{franson,singlephoton}.  Furthermore, they have also been used within quantum key distribution \cite{etime,largealpha,fransonqkd}.  In particular, energy-time entanglement can be used to encode multiple bits per photon pair \cite{brougham}.

While optical time-bin encoded states have been generated in experiments, performing suitable measurement on these states can be challenging.  In particular, it is difficult to realize experimentally measurements within a pair of MUBs, when the dimensions of the state space is large.  Measuring within the time of arrival basis can easily be achieved, provided the detectors can resolve the width of the time bins.  However, measuring in a basis that is mutually unbiased, with respect to the arrival time, can be very difficult.  One approach, that works for systems of dimension $d=2^M$, is to us a linear optical network of Franson interferometers \cite{fransonqkd}.  This requires one to align $d-1$ interferometers, which would be challenging even for small values of $d$.

In this paper we will outline a simple and compact experimental setup to measure in a superposition of several time bins.  The approach is based on a modified Mach-Zehnder interferometer, which acts like a cavity.  This enables us to make a measurement that approximates a true time-bin based MUB.  This setup can serve as a bases for many different time-bin based, high-dimensional quantum information protocols.  %The outline of the paper is as follows. In section \ref{sec2} we describe the experimental setup and explain how it allows one to measure within to MUBs.  The errors caused by the approximate nature of the MUBs is studied in section 3.  Finally, we discuss the results in section 4.

\begin{figure}
\center{\includegraphics[width=8.5cm,height=!]
{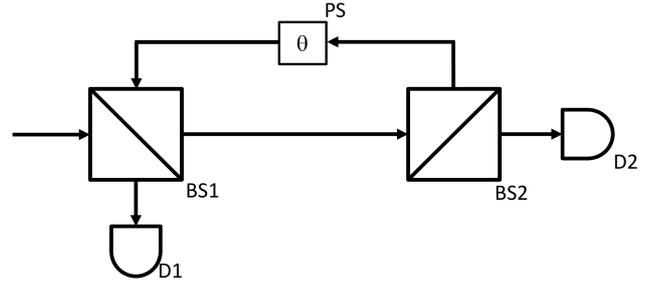}}
\caption{A diagram of the experimental setup.  BS1 and BS2 are both highly reflecting beamsplitters and PS is a phase shifter that gives a phase shift of $\theta$.  The total phase shift for one round trip of the interferometer is thus $\phi=\theta+\pi$.  Detection at D2 corresponds to a security check, while detection at D1 gives the time of arrival information and hence the key bits.}
\label{fig1}
\end{figure}

{\bf Using a cavity to measure in two MUBs}. - Suppose we prepare a photon in a state with uncertainty in it's time of arrival.  The arrival time of the photon will be divided into $d$ time slots, which we label from 1 to $d$.  Let $|n\rangle$ represent the state corresponding to a photon being in the $n$-th time bin.  Measuring the photon's time of arrival will be equivalent to projecting onto the basis $\{|n\rangle\}$.   We now seek another basis, which is mutually unbiased with respect to $\{|n\rangle\}$.  One such MUB is 
\begin{equation}
\label{mubs}
|\varphi_k\rangle=\frac{1}{\sqrt{d}}\sum_{n=0}^{d-1}{\exp\left(\frac{2\pi ink}{d}\right)|d-n\rangle},\;\;k=0,1,...,d-1.
\end{equation}
It can easily be verified that $|\langle m|\varphi_n\rangle|^2=1/d$.  One approach to implementing a measurement within the basis $|\varphi_k\rangle$, is to use an optical network \cite{fransonqkd,opticalnetwork}.  However, this approach requires one to align several interferometers, which is very difficult for large $d$.  The problem with making measurements in a superposition of time bins is that we requre each photon amplitude to interfere with other, temporally separated amplitudes.  A simple way of achieving this would be to use a cavity.  By carefully designing the cavity, one can use this interference to construct a measurement that is a very good approximation to projecting onto the state $|\varphi_k\rangle$.  The setup we use is a modified Mach-Zehnder interferometer, as shown in figure \ref{fig1}.  The two beam-splitters are both chosen to be highly reflective.  The phase shifter imparts a phase shift of $\theta$.  In addition to this, reflections at the two beam-splitters will also give a phase shift.  For a single round trip of the cavity, a photon (or pulse) would pickup a phase shift of $\phi=\theta+\pi$.

To understand how the setup allows us to measure within a MUB, it is helpful to consider the action on a single photon input state of the form $|\xi\rangle=\sum_k{e^{i\gamma_k}|d-k\rangle}$.  Suppose we obtain a click at D2 within the $N$-th time bin, where $N\ge d$.  The click could correspond to a photon within the $d$-th input slot, which would have taken $N-d$ round trips of the cavity before exiting.  A simple consequence of having taken $N-d$ round trips is that the photon will have acquired a phase of $\exp{i(N-d)\phi}$.  There is also a component corresponding to a photon that originated within the $d-1$ time slot.  This would have taken $N-d+1$ round trips and acquired a phase of $\exp{i(N-d+1)}$, before exiting the cavity.  Similarly, each of the $d$ time bins could have been the origin of the photon that was detected.  A simple calculation shows that detecting a photon at D2, within a time slot $N$ ($N\ge d$), can be thought of as projecting onto 
\begin{eqnarray}
\label{retrostate}
|\Gamma_N(\phi)\rangle=|T_1||T_2|(|R_1||R_2|)^{N-d}e^{i(N-d)\phi}\nonumber\\
\times\left[\sum_{n=0}^{d-1}{(|R_1||R_2|)^ne^{in\phi}|d-n\rangle}\right],
\end{eqnarray}
where $T_1$, $R_1$, $T_2$ and $R_2$ are the transmission and reflection coefficients for beam-splitters one and two, respectively.  If we set $\phi=2\pi k/d$ and make $|R_1|$ and $|R_2|$ close to one, then this state approximates $|\varphi_k\rangle$.  %The prefactor $(|R_1||R_2|)^{N-d}$ takes account of the fact that the probability of obtaining a click decreases with time.  This is a simple consequence of the fact that we are more likely to detect the photon at earlier times.  
Obtaining a click within {\it any} time bin, $N\ge d$, corresponds to projecting onto a state that approximates $|\varphi_k\rangle$.  If we obtain a click at D2 within a time slot before the $d$-th one, then we do not project onto the desired state as all $d$ components have not full entered the cavity.  

The average time we wait for a detection at D2 will increase as $|R_1|$ and $|R_2|$ get closer to one.  In particular, in the limit of $|R_1|$ and $|R_2|\rightarrow 1$, the average time we must wait will tend to infinity.  In turn, this means that the probability to detect a photon at D2 will go to zero.  

One point we must consider is that the cavity can only preserve the coherence between the photon amplitudes for a finite period of time.  It is thus necessary to impose an upper limit for when a click at D2 projects onto (\ref{retrostate}).  Let $N'$ be the last acceptable time slot.  The maximum possible value for $N'$ will depend on the $Q$ factor of the cavity and thus will increase as $|R_1|$ and $|R_2|$ increase \cite{laserbook}.  However, as we will explain in section 3, the effects of certain errors can be decreased by discarding detection events that occur at later times.  This procedure is equivalent to decreasing the value for $N'$.  We thus find that we project onto (\ref{retrostate}) when ever we get a click at D2 within a time bin $N$, where $d\le N\le N'$.

Controlling the value of the phase shifter allows one to select different values for the total phase shift $\phi$.  By choosing $\phi=2\pi k/d$, we can approximately project onto $|\varphi_k\rangle$.  It is thus possible to approximately project onto any of the basis states $|\varphi_k\rangle$, by simply changing the phase shifter.  One can thus obtain the full measurement statistics for the basis $\{|\varphi_k\rangle\}$.
  
The setup also allows one to measure within the basis $\{|n\rangle\}$.  If we detect a photon at D1, within the time slots $1$ to $d$, then this will correspond to measuring within the time of arrival basis.  We thus see that the setup shown in figure \ref{fig1} provides a compact way of effectively measuring within two MUBs.

Thus far we have not discussed the spectral widths of the time-binned photons.  This will be related to the size of the time bins, which in turn effects the path length of the cavity.  In particular, the smaller the temporal width of the time bins, the smaller we must choose the cavity's path length.  If we choose the time-bin's temporal width to be as small as possible, then we increase the dimensions of our system.  Furthermore, this should also ensures that the free spectral range is sufficiently large so that we need not worry about the effects of spectral filtering \cite{laserbook}.

{\bf The accuracy of the approximate MUBs}. - The scheme outlined previously, allowed one to approximately project onto anyone of the states $|\varphi_k\rangle$.  For this to be of practical use, it is vital to determine how good the approximation is.  One way of achieving this is to determine how well we could discriminate between the basis states $|\varphi_k\rangle$.  The idea is as follows.  Alice will prepare a photon in one of the orthogonal states $|\varphi_k\rangle$.  She will then send this state to Bob who will input it to the cavity described in figure 1.  By setting his phase to an appropriate value, he can approximately project onto one of the basis states.  If the cavity allowed one to implement the measurement perfectly, then Bob would be able to determine whether Alice had sent the basis state corresponding to his phase setting.  For example, if Alice sends $|\varphi_k\rangle$ and Bob has set $\phi=2\pi m/d$, where $m\ne k$, then Bob should never obtain a click at D2, within the correct time window.  However, the approximate nature of his measurement means that he will sometimes see a click.  This would correspond to an error.  The probability of error, given that Alice prepares the state $|\varphi_k\rangle$ and Bob uses $\phi=2\pi m/d$, will be
\begin{equation}
P(m|k)=\sum_{N=d}^{N'}{|\langle \Gamma_N\left(\frac{2\pi m}{d}\right)|\varphi_k\rangle|^2}.
\end{equation}
Care must be taken with this probability as $P(m|k)$ can sometimes be small due to the fact that we have a large probability to measure within the time-of-arrival basis.  To avoid this problem, we define the total error as
\begin{equation}
\label{errortot}
P_E=\frac{\sum_{m\ne k}^d{P(m|k)}}{\sum_{n=1}^d{P(n|k)}}.
\end{equation}
From the symmetry in equation (\ref{errortot}), it is clear that $P_E$ does not depend on which state Alice prepares.  Without loss of generality, we assume that Alice prepares the state $|\varphi_0\rangle$.  Figure 2 shows a plot of $P_E$ as a function of $|R|^2=|R_1|^2=|R_2|^2$, for different values of $d$.  It can be seen that as $|R|^2$ increases, the error decreases.  This leads to the important result: one can make $P_E$ arbitrarily small by choosing the reflectivities sufficiently large.  Figure 2 also shows $P_{D2}$, the total detection probability at D2.  We see that as $|R|^2$ increases, $P_{D2}$, and hence also the count rate, decreases.

\begin{figure}
\center{\includegraphics[width=8cm,height=!]
{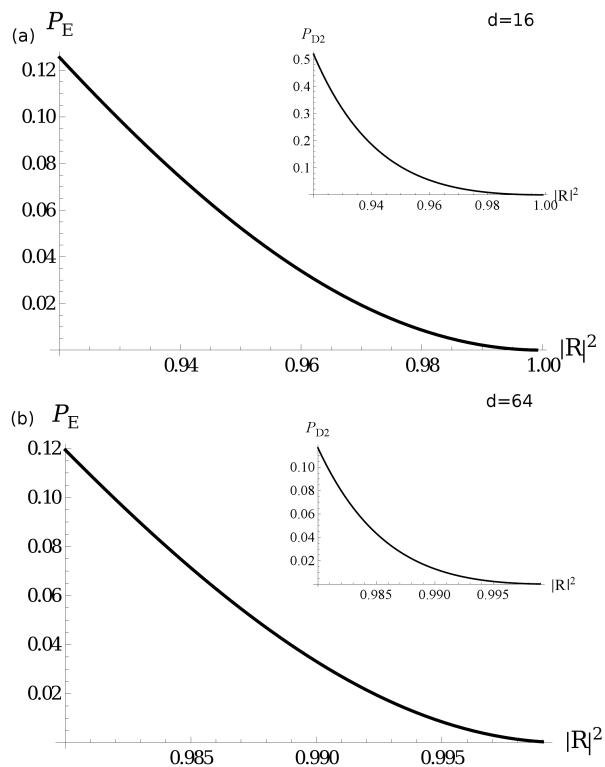}}
\caption{A plot of $P_E$ and $P_{D2}$, the total detection probability at D2, against $|R|^2=|R_1|^2=|R_2|^2$, for two different values of $d$, the number of time bins.  Figure (a) is for 16 time bins, {\it i.e.} $d=16$, while figure (b) is for 64 time bins.  }
\label{fig2}
\end{figure}

Any experimental implementation of the protocol will have additional errors associated with imperfections in the setup.  One important source of errors is misalignment of the interferometer.  For example, the length of the cavity might not be exactly equal to the spacing of the incoming time bins.  This would result in a mismatch between the incoming photon amplitude and those within the cavity.  This would `wash out' the intended interference effect.  If the path mismatch is too large, then the interference will be completely lost.  For small mismatch, the effects will be to increase the error, $P_E$.  It can be shown that the effect of a small mismatch can be modeled by introducing effective reflectivities, which are less than the actual ones.  To be more precise, if the reflectivities where $|R|^2=|R_{1,2}|^2$, then the effect of a small miss match will be to yield an error $P_E$ that corresponds to the effective reflectivity $|R'|^2<|R|^2$.  The effects of path mismatch can thus be counteracted by increasing the actual reflectivity of the beam-splitters.

Path mismatch is not the only source of errors.  The exact nature and importance of the other errors will depend on the particular application.   We find, however, that most errors can be reduced by increasing the values of the reflectivities.  One exception to this are errors due to dark counts.  The effects of dark count induced errors can, however, be minimised by reducing the cut-off $N'$.  The idea behind this is that the longer we have to wait for a detection, the greater the chance of it being due to a dark count.  The price we pay for reducing the error, is that we will have less security checks and will thus need to collect more data.  There is thus a trade-off between the count rate and the error rate.  A more detailed analysis of the errors will be presented elsewhere.

{\bf Conclusions}. - A simple and experimentally achievable means of creating high-dimensional states is to encode in the arrival time of a photon.  One can obtain a $d$-level system by simply dividing the arrival time into $d$ time slots.  However, if we are to make use of these states then we must be able to effectively control them.  A basic task for any application, is to extract information by making measurements.  Measuring within the time of arrival basis is straightforward.  Making measurements in a MUBs, such as (\ref{mubs}), can however, be very challenging.  We have described a compact scheme that, with high accuracy, allows one to approximately measure within two MUBs.  The approach was to use a modified Mach-Zehnder interferometer, which operated like a cavity.  This allowed photon amplitudes within various different time bins to interfere with each other.  By controlling a phase shifter, we can tailor the interference so as to approximately project onto any one of desired MUB states.

The states we project onto, are approximations of the desired ones.  The non-ideal nature of these states was investigated in terms of the error one would have if one wanted to discriminate between the true MUB states.  It was found that the error could be made arbitrarily small by increasing the values of the reflectivities of the first and second beam-splitters.  However, increasing the reflectivities decreases the probability of the photon being detected within an appropriate time slot, which would enact the measurement in the MUB.  There is thus a trade off between decreasing the error and the probability of making the measurement.  Nevertheless, it was shown that the error can be made sufficiently small, for reasonable values of the reflectivities.

The scheme we have outlined has applications with regards to experiments on high-dimensional photonic states.  For example, it can be used as a basis for tests of the non-locality of high-dimensional energy-time entangled states.  Similarly, the setup has applications within the field of quantum information.  One obvious example is in time-bin based high-dimensional QKD.  This approach offers the promise of encoding multiple key bits on each photon.  In particular, entanglement based protocols have been shown to allow for 10 bits (unsecured) to be encoded on each photon pair, under realistic but challenging experimental conditions \cite{brougham}.

\acknowledgments
We thank E. Eleftheriadou, N. L{\"u}tkenhaus and D. J. Gauthier for useful discussions.  This research was supported by the DARPA InPho program through the US Army Research Office award W911NF-10-0395.  %SMB also acknowledges the Royal Society and the Wolfson Foundation for support.

\end{document}